\documentclass[12pt]{article}
\pdfoutput=1
\usepackage{jhep-mod}
\usepackage{bm}
\usepackage{amssymb,amsmath,amsthm}
\usepackage{mathrsfs} 
\usepackage{graphics}
\DeclareMathOperator{\tr}{tr}

\def\widebar{\overline}

\def\x{{\mathbf{x}}}

\def\0{{\mathbf{0}}}

\def\TODAY{Thursday 21 June 2012}
\begin{document}
\title{Gordon and Kerr--Schild ans\"atze in massive and bimetric gravity}
\author{Valentina Baccetti, Prado Martin-Moruno, {\rm and} Matt Visser}
\affiliation{
School of Mathematics, Statistics, and Operations Research,\\
Victoria University of Wellington, PO Box 600, Wellington 6140, New Zealand}
\emailAdd{valentina.baccetti@msor.vuw.ac.nz}
\emailAdd{prado@msor.vuw.ac.nz}
\emailAdd{matt.visser@msor.vuw.ac.nz}
\abstract{
We develop the ``generalized Gordon ansatz'' for the ghost-free versions of both massive and bimetric gravity, an ansatz which is general enough to include almost all  spacetimes commonly 
considered to be physically interesting,  and restricted enough to greatly simplify calculations. The ansatz allows explicit calculation of the matrix square root 
$\gamma = \sqrt{g^{-1} f}$ appearing as a central feature of the ghost-free analysis. 
In particular, this ansatz automatically allows  us to write the effective stress-energy tensor as that corresponding to a perfect fluid. 
A qualitatively similar ``generalized Kerr--Schild ansatz'' can also be easily considered, now leading to an effective stress-energy tensor that corresponds to a null fluid. 
Cosmological implications are considered, as are consequences for black hole physics. Finally we have a few words to say concerning the null energy condition in the framework provided by these ans\"atze.


\bigskip
\noindent
\TODAY;  \LaTeX-ed \today.
}
\maketitle


\clearpage
\section{Introduction}

The ghost-free massive and bimetric gravities have recently attracted considerable attention. Early (pre-ghost-free) work
included~\cite{Isham:1971gm, Boulware:1973my, Visser:1997}, while notable contributions to the recent ghost-free discussion
include~\cite{Hassan:2011zd, deRham:2010ik, deRham:2010kj, Hassan:2011vm, Hassan:2011hr, Hassan:2011tf, Hassan:2011ea,  Hassan:2012qv, Hinterbichler:2012cn}. 
Recent work on the relation between massive gravity and bimetric gravity has been reported in~\cite{Baccetti:2012bk}, and the status of the
null energy condition (NEC) is explored in~\cite{Baccetti:2012zz}. Cosmological implications are discussed 
in~\cite{deRham:2011by, Chamseddine:2011bu, D'Amico:2011jj, Gumrukcuoglu:2011ew, vonStrauss:2011mq, Volkov:2011an, Comelli:2011zm}, 
while the status of black holes is considered in~\cite{Koyama:2011xz, Koyama:2011yg, Berezhiani:2011mt, Sbisa:2012zk, Comelli:2011wq, Volkov:2012wp, Deffayet:2011rh, Banados:2011hk}. 

An overarching theme of all the ghost-free technical calculations is the need for manipulating and working with the 
matrix square root $\gamma = \sqrt{g^{-1} f}$ built out of the foreground and background metrics. Some particularly useful 
mathematical results related to dealing with this quantity are reported in~\cite{Baccetti:2012zz}. In the current article we shall
investigate a specific ansatz for the relation between foreground and background metrics that is sufficiently general to be able to
represent almost all spacetimes of physical interest, while being simultaneously sufficiently special to make computations simple. 
The ansatz is inspired by the ``analogue spacetime'' programme~\cite{survey, LRR, ergoregion}, but has wide applicability and validity
outside of that original context. (Note that the analogue spacetime  programme could play a relevant role in this context not only as an 
inspiration, but even as a scenario where bimetric models could naturally emerge~\cite{Jannes:2011em}.)

\smallskip
The paper is organized as follows:
The general framework for ghost-free massive and bimetric gravity is very briefly summarized in section~\ref{framework}. 
Then in section~\ref{ansatz1} we introduce a particularly powerful ansatz which can be obtained by generalizing the Gordon metric 
(normally arising  in relativistic optics).
This ``generalized Gordon ansatz'' allows us to write the effective stress-energy  tensor as that of a perfect fluid, 
see subsection~\ref{perfectfluid}. 
Despite the rather strong conclusions implied by this ansatz, it is still general enough to describe the 
most common physically interesting spacetimes considered in the literature --- see in particular subsection~\ref{general}. 
In section~\ref{s:ks} we present the ``generalized Kerr--Schild ansatz'' and discuss the implied stress-energy tensor, now that of a null fluid.
We conclude with a discussion in section~\ref{discussion}.

\section{Stress energy tensor in bimetric gravity}\label{framework}

The action of bimetric gravity can be expressed quite generally as~\cite{Baccetti:2012bk, Baccetti:2012zz}
\begin{eqnarray}\label{actionbg}
S&=&-\frac{1}{16\pi G}\int d^4x\sqrt{-g} \left\{R(g) + 2\,\Lambda-2\,m^2 L_\mathrm{int}(g, f)\right\} +S_{({\rm m})}
\nonumber\\
&&
-\frac{\kappa}{16\pi G}\int d^4x\sqrt{-f} \left\{\widebar{R} (f)  +2\, \widebar\Lambda \right\} +\epsilon\,\widebar S_{({\rm m})}.
\end{eqnarray}
Here $S_{({\rm m})}$ and $\widebar S_{({\rm m})}$ are the usual matter actions, with the foreground and background matter fields being 
coupled only to the foreground and background metrics $g_{\mu\nu}$ and $f_{\mu\nu}$, respectively. All interactions between these two sectors are confined to the term $ L_\mathrm{int}(g, f)$, which is an algebraic function of $g$ and $f$.
We can recover the action of massive gravity by considering $\kappa=\epsilon=0$~\cite{Baccetti:2012bk}. In this case,
we would have an aether theory in which the dynamics of the physical metric $g_{\mu\nu}$ depends on a now non-dynamical background metric $f_{\mu\nu}$.
(Up to this point,  the analysis is identical to the pre-ghost-free analyses of the 1970s~\cite{Isham:1971gm, Boulware:1973my}, and later~\cite{Visser:1997}.)

As is now well-known~\cite{Hassan:2011zd}, the action (\ref{actionbg}) is ghost-free if and only if the interaction term can be written as a linear
combination of the elementary symmetric polynomials of the eigenvalues of the  matrix $\gamma=\sqrt{g^{-1} f}$.
In view of the results of~\cite{Baccetti:2012zz} we can write
\begin{equation}
f_{\mu\nu} = \gamma^\sigma{}_\mu \; g_{\sigma\rho} \; \gamma^\rho{}_\nu; \qquad\hbox{that is} \qquad f = \gamma^T \; g \; \gamma.
\end{equation}
We can then, in 3+1 dimensions, express the interaction Lagrangian as~\cite{Baccetti:2012bk}
\begin{equation}\label{Lint}
 L_\mathrm{int}=\alpha_1\,e_1(\gamma)+\alpha_2\,e_2(\gamma)+\alpha_3\,e_3(\gamma),
\end{equation}
where as usual
\begin{eqnarray}\label{symm}
 e_1(\gamma) &=&\tr[\gamma];\\
e_2(\gamma) &=&\frac{1}{2}\left(\tr[\gamma]^2-\tr[\gamma^2]\right);\\
e_3(\gamma) &=&\frac{1}{6}\left(\tr[\gamma]^3-3\tr[\gamma]\tr[\gamma^2]+2\tr[\gamma^3]\right).
\end{eqnarray}
The two remaining non-vanishing polynomials, $e_0(\gamma)=1$ and $e_4(\gamma)=\det(\gamma)$, have been absorbed into the foreground and background 
cosmological constants.
The three parameters appearing in the interaction term are not fully independent. If one additionally requires that $L_\mathrm{int}$ should (for weak fields)
 be of the canonical Fierz--Pauli form, then~\cite{Hassan:2011vm, Baccetti:2012bk}
\begin{equation}\label{relation}
 \alpha_1+2\alpha_2+\alpha_3=-1.
\end{equation}
Both metrics have exactly the same status. The interaction term satisfies the reciprocity relation~\cite{Hassan:2011zd, Baccetti:2012bk}
\begin{equation}\label{intbg}
\sqrt{-g} \; L_\mathrm{int}(\gamma) = \sqrt{-g} \; \sum_{i=0}^4 \alpha_i \; e_i(\gamma) =  \sqrt{-f} \; \sum_{i=0}^4 \alpha_{4-i} \; e_i(\gamma^{-1})
=\sqrt{-f} \; \widebar L_\mathrm{int}(\gamma^{-1}) .
\end{equation}
That is, the entire theory could be equivalently re-expressed using $f$ as the foreground metric and $g$ as the background.
Varying the action (\ref{actionbg}) with respect the two metrics, we obtain two sets of equations of motion~\cite{Baccetti:2012bk}:
\begin{equation}\label{motiong}
 G^{\mu}{}_{\nu}-\Lambda \,\delta^\mu{}_\nu =m^2\,T^{\mu}{}_{\nu}+8\pi G \;T^{({\rm m})\mu}{}_{\nu},
\end{equation}
and
\begin{equation}\label{motionf}
 \kappa\,\left(\widebar{G}^{\mu}{}_{\nu}-\widebar\Lambda\,\delta^\mu{}_\nu \right) =
m^2\, \widebar{T}^{\mu}{}_{\nu} +\epsilon\,8\pi G \,\widebar T^{({\rm m})\mu}{}_{\nu}.
\end{equation}
Here
\begin{equation}\label{Tg}
 T^{\mu}{}_{\nu}=\tau^{\mu}{}_{\nu}-\delta^{\mu}{}_{\nu}\,L_\mathrm{int},
\end{equation}
and
\begin{equation}\label{Tf}
 \widebar{T}^{\mu}{}_{\nu}=-\frac{\sqrt{-g}}{\sqrt{-f}}\;\tau^{\mu}{}_{\nu},
\end{equation}
with 
\begin{equation}\label{taudef}
 \tau^{\mu}{}_{\nu}=\gamma^\mu{}_\rho\,\frac{\partial L_\mathrm{int}}{\partial \gamma^{\nu}{}_{\rho}}.
\end{equation}
The indices of equation (\ref{motiong}) and (\ref{motionf}) must be raised and lowered using $g$ and $f$, respectively. The equations of motion of the $g$-space ($f$-space) are modified with respect to those
of general relativity by the introduction of an \emph{effective stress-energy tensor} associated to the interaction between the two geometries (and the quantity $\kappa/\epsilon$). 
It is a slightly non-trivial exercise, see~\cite{Baccetti:2012zz}, to verify that the effective stress-energy tensors are indeed symmetric.
In massive gravity with a non-dynamical background one would only have the first set of equation of motions
(\ref{motiong}), see~\cite{vonStrauss:2011mq, Baccetti:2012bk}.
Both effective stress-energy  tensors fulfill the Bianchi-inspired constraints
\begin{equation}\label{Bianchi}
\nabla_\mu T^{\mu}{}_{\nu}=0; \qquad \widebar{\nabla}_\mu \widebar{T}^{\mu}{}_{\nu}=0.
\end{equation}
Once one constraint is enforced, for example $\nabla_\mu T^{\mu}{}_{\nu}=0$, then the other
is also automatically fulfilled~\cite{vonStrauss:2011mq, Volkov:2011an}.

The modifications to the two sets of equations of motion are very closely related. Everything can be expressed in terms of  a single mixed-index tensor $ \tau^{\mu}{}_{\nu}$, 
 which can be written as a cubic polynomial in the matrix $\gamma$. Specifically~\cite{Baccetti:2012zz}
\begin{eqnarray}\label{stress}
 \tau^{\mu}{}_{\nu}=
 \left(\alpha_1+\alpha_2\,e_1(\gamma)+\alpha_3\,e_2(\gamma)\right) \gamma^\mu{}_\nu -
\left(\alpha_2+\alpha_3\,e_1(\gamma)\right) \{\gamma^2\}^\mu{}_\nu +
\alpha_3 \{\gamma^3\}^\mu{}_\nu .
 \end{eqnarray}
This yields the effective stress-energy  tensor as viewed by the $g$-space, up to a cosmological constant contribution coming from second term
in the RHS of equations (\ref{Tg}). Explicitly
\begin{eqnarray}\label{stressT}
 T^{\mu}{}_{\nu}=
 -\left(\alpha_1\,e_1(\gamma)+\alpha_2\,e_2(\gamma)+\alpha_3\,e_3(\gamma)\right)\; \delta^\mu{}_\nu + \tau^{\mu}{}_{\nu}.
 \end{eqnarray}
 We also note the foreground-background symmetry
\begin{equation}\label{tensors}
\sqrt{-g} \; T^{\mu}{}_{\nu} + \sqrt{-f} \; \widebar T^{\mu}{}_{\nu} = - \sqrt{-g} \; L_\mathrm{int}(\gamma) \;\delta^{\mu}{}_{\nu}  = -\sqrt{-f} \; \widebar L_\mathrm{int}(\gamma^{-1})\;  \delta^{\mu}{}_{\nu} .
\end{equation}

\section{Generalized Gordon ansatz}\label{ansatz1}
Let us now consider the following ansatz relating the two metrics:
\begin{equation}\label{ansatz-a}
f_{\mu\nu} = \Omega^2 \{ g_{\mu\nu} + \xi \;V_\mu V_\nu  \}.
\end{equation}
Here $V^\mu$ is a unit timelike vector with respect to $g_{\mu\nu}$, so that $g_{\mu\nu}V^\mu V^\nu=-1$, and we set $V_\mu = g_{\mu\nu} V^\nu$. Furthermore $\Omega$ and $\xi$ are arbitrary functions.
Noting that both metrics must have Lorentzian signature, we see $\xi<1$.  So we can write $\xi = 1 - \zeta^2$,  with $\zeta\neq0$.  Furthermore, without any loss of generality we can choose $\zeta$ to be strictly positive, that is $\zeta>0$.
Then
\begin{equation}\label{ansatz-b}
f_{\mu\nu} =  \Omega^2 \{ (g_{\mu\nu}  +V_\mu V_\nu )  - \zeta^2 \;V_\mu V_\nu \}.
\end{equation}
Mathematically, this ansatz can be thought of a conformal transformation $\Omega$, combined with a stretch $\zeta$ along the timelike direction parallel to $V$. 
\begin{figure}[htbp!]
\begin{center}
\includegraphics[scale=0.50]{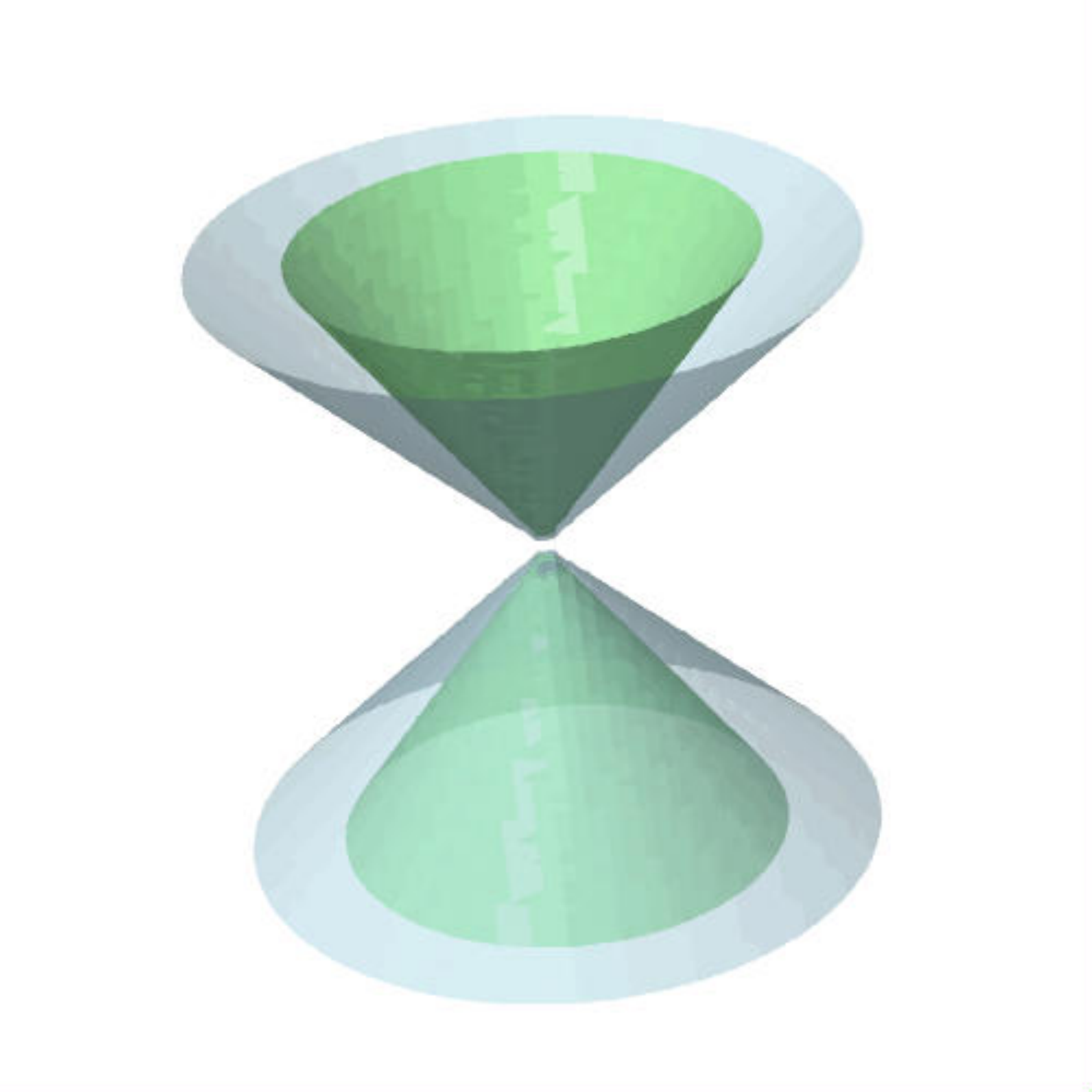}
\caption{Light cones for the generalized Gordon ansatz: Depending on the value of the parameter $\zeta$ the foreground light cones lie strictly inside, on top of, or strictly outside the background light cones.}
\label{F:Gordon}
\end{center}
\end{figure}
Note that (see figure~\ref{F:Gordon}) the relative position of the light cones is very strongly correlated.
\begin{itemize}
\item For $0<\zeta<1$ the light cones of $f$ lie strictly inside the light cones of $g$.
\item For $\zeta=1$ the light cones of $f$ and $g$ coincide; the two metrics are conformally related.
\item For $\zeta>1$ the light cones of $f$ lie strictly outside the light cones of $g$.
\end{itemize}

\subsection{Historical background}\label{history}

Physically, this ansatz is a significant generalization of Gordon's 1923 metric, which was originally developed within the context of geometric optics~\cite{Gordon}; see particularly~\cite{survey, LRR} for recent discussions.
The original Gordon metric, which describes ray optics in an optical medium of 4-velocity $V^\mu$ and constant
refractive index $n$ in Minkowski space, is  expressed as
\begin{equation}\label{Gordon}
 \mathcal{G}_{\mu\nu}=(\eta_{\mu\nu}  +V_\mu V_\nu )  - \frac{1}{n^2} \; V_\mu V_\nu ,
\end{equation}
where $V_\mu=\eta_{\mu\nu}V^\nu $. Our ansatz (\ref{ansatz-b}) corresponds to a generalization of (\ref{Gordon}). One should:
\begin{itemize}
\item[(i)] Generalize the refractive index $n$ to be a position-dependent function ($1/n^2 \to \zeta^2$);
\item[(ii)] Introduce a position-dependent conformal factor $\Omega$, (which does not affect the light cone structure);
\item[(iii)] Consider a non-trivial input metric ($g_{\mu\nu}$ instead of $\eta_{\mu\nu}$);
\item[(iv)] Relabel the output metric  ($f_{\mu\nu}$ instead of $\mathcal{G}_{\mu\nu}$).
\end{itemize}
For this reason, we call our ansatz the ``generalized Gordon ansatz''. The ansatz can also naturally be inverted to express $g_{\mu\nu}$ in terms of $f_{\mu\nu}$:
\begin{equation}\label{ansatz-bb}
g_{\mu\nu} =  \Omega^{-2} \{ (f_{\mu\nu}  +\widebar V_\mu \widebar V_\nu )  - \zeta^{-2} \; \widebar V_\mu \widebar V_\nu \},
\end{equation}
where $\widebar V$ is now normalized in terms of $f_{\mu\nu} \widebar V^\mu \widebar V^\nu=-1$, and we set 
$\widebar V_\mu = f_{\mu\nu} \widebar V^\nu$. For consistency we must then enforce $\widebar V_\mu = \Omega\zeta \; V_\mu$ and $\widebar V^\mu = \Omega^{-1}\zeta^{-1} \; V^\mu$.

This  generalized Gordon ansatz also has exactly the same structure as the (geometric) relativistic acoustic metric of~\cite{Moncrief, Vikman, Visser:2010xv}; see particularly~\cite{survey, LRR} for recent discussions. 
Our ansatz, as expressed in equation (\ref{ansatz-b}),  exactly agrees with the form of the geometric acoustic metric
appearing in reference~\cite{Moncrief, Vikman, Visser:2010xv}. (Here we have denoted the second metric $f_{\mu\nu}$ instead of $\mathcal{G}_{\mu\nu}$.
Furthermore in the present situation $\zeta^2$ is not necessarily restricted to be less than one. In contrast, in~\cite{Moncrief, Visser:2010xv} $\zeta^2=c_s^2/c^2\leq 1$, simply because the speed of sound is subluminal, though ``tachyacoustic cosmologies'', based on~\cite{Vikman}, are now fashionable.)

\subsection{Perfect fluid effective stress-energy  tensor}\label{perfectfluid}
Using this ansatz we can express the matrix $\gamma$ in a simple form by noting that
\begin{equation}\label{gamma2}
g^{\mu\sigma} f_{\sigma\nu} = \Omega^2\{ (\delta^\mu{}_\nu   + V^\mu V_\nu  ) - \zeta^2 \; V^\mu V_\nu  \}=(\gamma^2)^\nu{}_\nu .
\end{equation}
Then the matrix square root is easy to extract
\begin{equation}\label{gamma3}
\gamma^\mu{}_\nu  = (\sqrt{g^{-1} f} )^\mu{}_\nu  = \Omega \{ (\delta^\mu{}_\nu   + V^\mu V_\nu ) - \zeta \; V^\mu V_\nu   \}.
\end{equation}
Furthermore
\begin{equation}\label{gamma3b}
(\gamma^n)^\mu{}_\nu   = \Omega^n \{ (\delta^\mu{}_\nu   + V^\mu V_\nu ) - \zeta^n \; V^\mu V_\nu   \}.
\end{equation}
The symmetric polynomials appearing in the interaction term (\ref{Lint}), given by equations (\ref{symm}), also take
on a particularly simple form. They are
\begin{eqnarray}\label{symmA}
e_1(\gamma)=  \Omega \{ 3 + \zeta   \};\qquad
e_2(\gamma) = 3 \Omega^2 \{1+\zeta\};\qquad
e_3(\gamma) = \Omega^3 \{ 1+3\zeta\}.
\end{eqnarray}
Note that these expressions are linear in $\zeta$. This can be traced back to the fact that only one of the eigenvalues of $\gamma$ is at all $\zeta$-dependent, and that in a linear fashion.
Taking into account expressions (\ref{gamma2}), (\ref{gamma3}), and (\ref{symmA}), we see that equations 
(\ref{stress}) and (\ref{stressT}), imply
that the stress-energy  tensor takes a perfect fluid form:
\begin{equation}\label{perfect}
T^\mu{}_\nu  = p \; (\delta^\mu{}_\nu   + V^\mu V_\nu ) +  \rho \; V^\mu V_\nu.
\end{equation}
Here we have defined
\begin{equation}\label{rho}
\rho = \Omega\left(3\alpha_1 +3\alpha_2 \Omega + \alpha_3 \Omega^2\right),
\end{equation}
and
\begin{equation}\label{p}
p = -\Omega \left[(2 \alpha_1 +\alpha_2 \Omega) + (\alpha_1 +2\alpha_2 \Omega +\alpha_3 \Omega^2) \zeta\right].
\end{equation}
Note that these expressions are also linear in $\zeta$, ultimately for the same reason as above.
It is also useful to write
\begin{equation}\label{perfectau}
\tau^\mu{}_\nu  = \hat{p}\; (\delta^\mu{}_\nu   + V^\mu V_\nu ) +  \hat{\rho} \; V^\mu V_\nu  , 
\end{equation}
defining
\begin{equation}\label{taurho}
\hat{\rho} = -\Omega\zeta\left(\alpha_1+3\alpha_2\Omega+3\alpha_3\Omega^2\right),
\end{equation}
and
\begin{equation}\label{taup}
\hat{p} = \Omega \left[\alpha_1+\alpha_2\Omega(2+\zeta)+\alpha_3\Omega^2(1+2\zeta)\right].
\end{equation}
Here we have used the fact that $T^\mu{}_\nu $ and $\tau^\mu{}_\nu $ only differ in a cosmological constant contribution. Moreover, we have
\begin{equation}\label{prho}
\rho + p = \hat{\rho}+\hat{p}  = \Omega  (\alpha_1 +2\alpha_2 \Omega +\alpha_3 \Omega^2) (1-\zeta).
\end{equation}
Note that $\zeta=1$ ($\xi=0$) corresponds to a cosmological constant, as should be expected, since in this case ansatz (\ref{ansatz-a}) 
reduces to a conformal relation between the two metrics.

Moreover, due to the Bianchi-inspired constraint (\ref{Bianchi}), the stress-energy  tensor not only takes the perfect
fluid form (\ref{perfect}), but also obeys the usual conservation equations. These are
\begin{equation}\label{con1}
V^\mu \nabla_\mu\rho+(\rho+p)\nabla^\mu V_\mu=0,
\end{equation}
and
\begin{equation}\label{con2}
(\rho+p)V^\mu \nabla_\mu V_\nu +(g_{\mu\nu}+V_\mu V_\nu )\nabla^\mu p=0.
\end{equation}
On the other hand, in view of equations (\ref{rho}) and (\ref{prho}), it can be noted that the effective fluid automatically fulfills
the additional constraint:
\begin{equation}\label{condadi}
\nabla_\mu\rho=3\; \frac{\rho+p}{1-\zeta}\; \frac{\nabla_\mu \Omega}{\Omega}. 
\end{equation}
From equations (\ref{con1}) and (\ref{condadi}) one can conclude that, once $\Omega$ and $V^a$ have been fixed in (\ref{ansatz-b}),
the following equation is automatically fulfilled
\begin{equation}
\zeta=\frac{\nabla_a\left(\Omega^3V^\mu \right)}{\Omega^3\;\nabla_\mu V^\mu}.
\end{equation}
Finally, 
the effective stress-energy tensor in the $f$-space (\ref{Tf}) also takes on a perfect fluid form
\begin{equation}\label{perfectf}
\widebar T^\mu{}_\nu  = \widebar p \; (\delta^\mu{}_\nu   + \widebar V^\mu \widebar V_\nu ) +  \widebar \rho\; \; \widebar V^\mu \,\widebar V_\nu.
\end{equation}
Now $\widebar V_\mu = \Omega\zeta \; V_\mu$, as defined in the previous section, is normalized in terms of 
$f_{\mu\nu}\widebar V^\mu \widebar V^\nu=-1$, and so
\begin{equation}\label{pf}
 \widebar p=-\frac{1}{\Omega^4\zeta }\; \hat p; \qquad \widebar \rho=-\frac{1}{\Omega^4\zeta}\; \hat \rho.
\end{equation}
(This last step could equivalently be obtained by noting $\sqrt{-f} = \Omega^4 \zeta\;\sqrt{-g}$.)
Perhaps more tellingly
\begin{equation}
\bar\rho+\bar p = -\frac{1}{\Omega^4\zeta}\; (\hat \rho+\hat p) =  -\frac{1}{\Omega^4\zeta}\; (\rho+p). 
\end{equation}
Thus, the sum of the effective energy density plus the pressure of the effective perfect fluid due to the modification of general 
relativity in the background space
has opposite sign to that characterizing the effective perfect fluid in the foreground space. (This is a special case of the more general NEC-violating behaviour discussed in~\cite{Baccetti:2012zz}. For general background on the NEC and its variants, see~\cite{Barcelo:2002bv, w-matter, Ford:1994bj, Ford:1999qv}.) Some of the possible
consequences of this sign flip are discussed in the particular case of cosmological solutions in section~\ref{general}.

\section{Generality of the ansatz}\label{general}

Up to now, we have shown that the ansatz expressed through (\ref{ansatz-a}) or, equivalently, through (\ref{ansatz-b}),
implies that the bimetric-induced effective stress-energy  tensor takes the form of a perfect fluid. Moreover, the ansatz constrains the effects of the modifications
of the theory in the two spaces so tightly that (apart from the ``trivial'' case of pure cosmological constant) the violation of the NEC condition in one space follows straightforwardly from 
the fulfillment of the NEC in the other space~\cite{Baccetti:2012zz}.
Nevertheless, the ansatz is general enough to describe many different physically relevant situations. Some particularly relevant examples are considered below.

\subsection{Cosmological solutions}

Let us now assume that the metric $g_{\mu\nu}$ takes the form of a FLRW spacetime in a certain coordinate patch. This is
\begin{equation}\label{cosmo-g}
ds_g^2 = - dt^2 + a(t)^2 \left[ {dr^2\over 1- k r^2} + r^2 \{d\theta^2 + \sin^2\theta\; d\phi^2\} \right].
\end{equation}
Application of the ansatz (\ref{ansatz-b}) leads to
\begin{equation}\label{cosmo-f}
ds_f^2 = - \Omega(t)^2\left\{\zeta(t)^2\; dt^2 + a(t)^2 \left[ {dr^2\over 1- k r^2} + r^2 \{d\theta^2 + \sin^2\theta\; d\phi^2\} \right]\right\}. 
\end{equation}
This metric can be written as
\begin{equation}\label{cosmo-f2}
ds_f^2 = - N(t)^2 \; dt^2 + b(t)^2 \left[ {dr^2\over 1- k r^2} + r^2 \{d\theta^2 + \sin^2\theta\; d\phi^2\} \right],
\end{equation}
where the scale factor and lapse function of metric $f$ are
\begin{equation}\label{scalef}
b(t) = \Omega(t) \,a(t),
\end{equation}
and
\begin{equation}\label{lapse}
 N(t)=\Omega(t)\,\zeta(t),
\end{equation}
respectively.
Thus, the assumption of the generalized Gordon ansatz (\ref{ansatz-b}), when considering a cosmological foreground metric, implies a cosmological background metric
with a different scale factor and cosmic time,
as  should be expected. The only restrictions coming from forcing that relation is that the two metrics should both be diagonal in the same
coordinate patch,  and have the same sign of spatial curvature $k$, conditions that certainly should be expected to be fulfilled by general cosmological 
solutions~\cite{Volkov:2011an, Comelli:2011zm}.

The conservation equations (\ref{con1}) take on the usual cosmological form.
Thus
\begin{equation}\label{conscosmol}
\dot \rho + 3 (\rho+p) {\dot a\over a} = 0.
\end{equation}
Taking into account equation (\ref{prho}) and the derivative of equation (\ref{rho}), this implies
\begin{equation}\label{zzz}
{\dot\Omega\over\Omega} = -  (1-\zeta) {\dot a\over a} .
\end{equation}
So, defining the Hubble parameters for both metrics in the usual way, $H_g=\dot{a}/a$ and $H_f=\dot{b}/b$, and taking into
account the definition of $b$ (\ref{scalef}), we see that the two Hubble parameters are very tightly intertwined 
\begin{equation}\label{relationH}
H_f = \zeta\;  H_g.
\end{equation}
This is equivalent to the relation obtained in reference~\cite{Comelli:2011zm} to characterize their ``branch two'' solutions, 
the general class of cosmological solutions in bimetric gravity.

Moreover, rewriting (\ref{zzz}) as
\begin{equation}
{\dot\Omega\over\Omega} + {\dot a\over a}  = \zeta {\dot a\over a},
\end{equation}
in view of (\ref{scalef}) we see
\begin{equation}
{\dot b\over b} =  \zeta \; {\dot a\over a},
\end{equation}
whence
\begin{equation}\label{zeta-a}
\zeta = {\dot b\over \dot a} \; {a\over b} = {d \ln b/dt \over d \ln a/dt}.
\end{equation}
Taking into account (\ref{lapse}) and (\ref{scalef}), this implies a lapse function
\begin{equation}\label{N-a}
N(t) = {\dot b\over \dot a}.
\end{equation}
So, in this case, one can write the metric of the background space (\ref{cosmo-f2}) using only one arbitrary function $b(t)$:
\begin{equation}\label{cosmo-f3}
ds_f^2 = - \left(\frac{\dot b}{\dot a}\right)^2 dt^2 + b(t)^2 \left[ {dr^2\over 1- k r^2} + r^2 \{d\theta^2 + \sin^2\theta\; d\phi^2\} \right].
\end{equation}
Moreover, from equations (\ref{scalef}) and (\ref{zeta-a}), we can express the energy density (\ref{rho}) and pressure (\ref{p})
of the effective stress
energy tensor in the foreground space as
\begin{equation}
 \rho=\frac{b}{a}\left(3\alpha_1+3\alpha_2\; \frac{b}{a}+\alpha_3\; \frac{b^2}{a^2}\right),
\end{equation}
and
\begin{equation}
 p=-\frac{b}{a}\left(2\alpha_1+\alpha_2\; \frac{b}{a}\right)-\left(\alpha_1+2\alpha_2\; \frac{b}{a}+\alpha_3\; \frac{b^2}{a^2}\right)\frac{\dot b}{\dot a},
\end{equation}
respectively. The $g$-space Bianchi-inspired conservation equation is then automatically satisfied for arbitrary $b(t)$.

Let us briefly consider the implications on the cosmological evolution of the $g$-space and $f$-space due to
the inter-twining relation between the NECs. The simplest possibility is to have both NECs satisfied, which in this cosmological setting implies that
the two cosmologies are conformally related, with the contribution to the effective
stress-energy  being an arbitrary cosmological constant that can be tuned at will. 
(It must be emphasized that, in principle, more general solutions are possible
for particular combinations of the parameters appearing in $L_{{\rm int}}$.)

Nevertheless, the case when the NECs are \emph{not} saturated is more interesting, because
either the foreground or background geometry has a NEC-violating effective stress-energy tensor.
Assuming that the pathologies
are relegated to the ``other'' gravitational sector, that is, in our own sector $\rho>0$ and $p+\rho>0$, then, 
in view of equations (\ref{rho}), (\ref{taurho}), (\ref{prho}) and (\ref{pf}), we have two possibilities:
\begin{enumerate}
\item[(i)] Either $\widebar \rho>0$ and $\widebar p+\widebar \rho<0$, which is phantom energy.
\item[(ii)] Or $\widebar \rho<0$ and $\widebar p+\widebar \rho<0$, which corresponds to an even weirder kind of cosmological fluid known as dual dark energy \cite{Yurov:2006we}.
\end{enumerate}
Regarding (i), as is well known, phantom energy can easily lead to a big rip \cite{Caldwell:2003vq}, big freeze 
\cite{BouhmadiLopez:2006fu, BouhmadiLopez:2007qb},
future singularities \cite{Cattoen:2005dx}, and other oddities~\cite{MolinaParis:1998tx, Hochberg:1998vm, epoch, epoch2}. 
The only possibility to avoid future doomsday in this scenario, under the assumption $p+\rho>0$, seems to be by requiring
$\widebar p+\widebar \rho\rightarrow0$ asymptotically (implying the same behaviour for $p+\rho$), leading to a cosmology
which approximates a de Sitter behaviour at late times.
In the second case, (ii), it must be noted that this effective dual dark energy is compatible with a Lorentzian geometry 
($3\,H_{f}^2=m^2\widebar\rho/\kappa+\widebar\Lambda-k/b^2>0$), a positive cosmological constant $\widebar\Lambda>0$ (with 
$\widebar\Lambda=\Lambda_f+\alpha_4$), 
or in a spatially hyperbolic universe, $k<0$. As  was studied in reference \cite{GonzalezDiaz:2007yb} for the case of a constant ratio
$\widebar w = \widebar p/\widebar\rho$, a universe
filled with dual dark energy and equipped with a positive cosmological constant 
(with $k=0$) is completely regular and the $\widebar\Lambda$-term dominates at late times.
Thus, in this case, the $f$-universe is asymptotically de Sitter at late times (for $\widebar\Lambda>0$ and any $k$).

On the other hand, in massive gravity we are free to choose any arbitrary function $b(t)$ since there are no equations of motion (\ref{motionf})
for the background metric. Therefore,
we can use the graviton mass terms to generate by hand an effective stress-energy  contributing to cosmological fluid 
that has any properties we desire --- the large scale evolution of our universe is no longer governed by the matter in our universe, 
but can instead be forced to dance to the tune determined by the arbitrarily specified background geometry 
(see reference \cite{vonStrauss:2011mq}, and for earlier [pre-ghost-free] speculations along these lines see~\cite{Visser:1997}).

\subsection{Massive-gravity black holes}

As we have already pointed out, the present ansatz can be motivated by generalizing the Gordon metric (and the relativistic acoustic metric) appearing in the field
of analogue spacetimes~\cite{survey, LRR}. Following this spirit, and in view of the achievements of that field,
one might expect this ansatz to be able to describe black hole geometries, at least when the metric describing the ``medium''
is the Minkowski metric. It must be noted that considering one of the metrics to be flat would implicitly restrict our attention to massive
gravity, where one of the metrics is non-dynamical. This is the theory that we are implicitly assuming in the present subsection.

Furthermore, let us note that a Minkowski medium in expression (\ref{Gordon}) implies $g_{\mu\nu}=\eta_{\mu\nu}$, so $g_{\mu\nu}$ cannot be the dynamical metric,
whereas it is $\mathcal{G}_{\mu\nu}=f_{\mu\nu}$ that is dynamical.
Usually in massive gravity the physical dynamical metric is denoted by $g_{\mu\nu}$ and the background metric is $f_{\mu\nu}$, but this is simply a
matter of notation. (Recall the symmetry of the interaction Lagrangian (\ref{intbg}) which leads to the graviton contribution to the matter Lagrangian in massive gravity.)
We maintain the notation of (\ref{Gordon}) for consistency with the analogue model programme. (In particular, 
all the results of this subsection would remain unchanged by adopting and specializing the ``inverse ansatz'' (\ref{ansatz-bb}), considering $g_{\mu\nu}$ to be the dynamical
metric and $f_{\mu\nu}=\eta_{\mu\nu}$ to be non-dynamical, and introducing an extra normalization factor in the four-velocity (\ref{velocity}); since, in that case,
it would be $V$ that was a timelike unit vector with respect to the curved metric, while $\widebar V$ would be a timelike unit vector with respect to the Minkowski metric.)

It should be pointed out that assuming a flat background for massive gravity is not too 
strong an assumption when dealing with black holes, (though it would certainly be an overly stringent assumption for massive-gravity cosmology), at least as long as  we are not presuming
that both metrics can be written in a diagonal form in the same coordinate patch~\cite{Deffayet:2011rh}. (See also~\cite{Banados:2011hk}.) In fact, to our knowledge,
massive-gravity  black holes (not bimetric gravity black holes) have been studied in the literature only by assuming flat backgrounds~\cite{Koyama:2011xz, Koyama:2011yg, Berezhiani:2011mt, Sbisa:2012zk}.

In massive gravity with a conformally flat background metric, our ansatz (\ref{ansatz-a})
can be written as
\begin{equation}\label{bh}
f_{\mu\nu} =  \Omega^2\{ \eta_{\mu\nu} + \xi V_\mu V_\nu \}.
\end{equation}
There are \emph{at least} two ways of proceeding.

\subsubsection{Isotropic coordinates}
Take $V^\mu = (1,\0)$, and set $U^2=\Omega^2(1-\xi)$. Then
\begin{equation}
ds^2 = - U^2 dt^2 + \Omega^2 \|d\x\|^2. 
\end{equation}
This puts the metric into isotropic coordinates --- and it is well known that this form of the metric is certainly sufficient 
(in principle) to deal with the Schwarzschild and Reissner--Nordstr\"om geometries, though the Kerr and Kerr--Newman geometries cannot 
be put in this form~\cite{Visser:2004}. However in this coordinate system $\gamma=\sqrt{g^{-1}f}$ is singular (in the matrix sense) as 
any horizon (corresponding to $U=0$) is approached. Thus once $L_\mathrm{int}$ is ``turned on'' diagonal coordinate systems of this type 
behave unpleasantly --- the stress tensor $T^\mu{}_\nu$ is then a singular matrix at the horizon, and other 
approaches might be more profitable. 

\subsubsection{Horizon-penetrating coordinates}
Let us now adopt horizon-penetrating coordinates. Choose $\Omega=1$, and express the background metric in the following gauge:
\begin{equation}\label{flatf}
ds_{\eta}^2 = \eta_{\mu\nu}dx^\mu dx^\nu=-dt^2 + dr^2 + r^2 \{d\theta^2 + \sin^2\theta\; d\phi^2\},
\end{equation}
Take the four-velocity to be
\begin{equation}\label{velocity}
V^\mu = \gamma\, ( 1,\,-\beta^i);  \qquad V_\mu = -\gamma\, (1,\,\beta^i); \qquad  \beta^i=(\beta^r,\,0,\,0).
\end{equation}
Note that the usual relation between $\beta$ and $\gamma$ is recovered when the normalization of $V_\mu$ with respect to $\eta_{\mu\nu}$
is imposed. (The $\gamma$ appearing here should not be confused with the matrix $\gamma = \sqrt{g^{-1} f}$ depending on the two metrics;
its form would be that of the usual relativistic quantity. The intended meaning should be clear from context.)
Thus, taking into account (\ref{bh}), (\ref{flatf}) and (\ref{velocity}), we can write
\begin{eqnarray}\label{bhgamma}
d s_f^2 = -\left(1-\xi\gamma^2\right) dt^2+ 2 \xi\gamma \sqrt{\gamma^2-1} dt dr 
+
\left[1+\xi\left(\gamma^2-1\right)\right] dr^2 +r^2 d\Omega_{(2)}^2.
\end{eqnarray}
Here $d\Omega_{(2)}^2=d\theta^2 + \sin^2\theta \; d\phi^2$ must not be confused with the conformal factor $\Omega$.

Let us now consider the specific function
\begin{equation}\label{gammar}
\gamma(r) = \sqrt{1+K/r}.
\end{equation}
Then the foreground metric (\ref{bhgamma}) can be written as
\begin{eqnarray}
d s_f^2 = -\left(1-\xi-{\xi K\over r}\right) dt^2 + 2\xi {\sqrt{K(K+r)}\over r} dt dr 
+ \left(1+{\xi K\over r}\right) dr^2 
+ r^2 d\Omega_{(2)}^2.
\quad
\end{eqnarray}
In its current form it is not canonically normalized at spatial infinity. If we now re-define the time variable as 
$\tilde{t}=\sqrt{1-\xi}\,t$, then
\begin{equation}
d s_f^2 = -\left(1-{\xi K\over (1-\xi) r}\right) d\tilde{t}^2 + 2 \xi {\sqrt{K(K+r)}\over \sqrt{1-\xi} \; r} d\tilde{t} dr + 
\left(1+{\xi K\over r}\right) dr^2 + r^2 d\Omega_{(2)}^2,
\end{equation}
which by explicit computation can easily be seen to be Ricci flat.
This allows us to identify 
\begin{equation}
2 m = {\xi K\over (1-\xi)};
\end{equation}
and so 
\begin{equation}\label{bhepsilon}
d s_f^2 = -\left(1-{2m\over r}\right) d\tilde{t}^2 + 2{\sqrt{2m[2m(1-\xi)+\xi r]}\over r} d\tilde{t} dr 
+ \left(1+{2m(1-\xi)\over r}\right) dr^2 + r^2 d\Omega_{(2)}^2.
\end{equation}
Now clearly in the limit $\xi\rightarrow1$  one recovers the Painlev\'e--Gullstrand form of the Schwarzschild metric,
whereas if $\xi=0$, then the black hole is expressed using the Eddington--Finkelstein metric. (For a related analysis, see~\cite{Rosquist:2003qs}.)

One might think that there is some problem in performing a time re-definition once the gauge freedom is completely used to fix the form
of the background metric (\ref{flatf}). However, the same result can be obtained in a slightly different manner without explicitly performing any coordinate
change. Let us start from the background metric already written using the new time coordinate. This is
\begin{equation}\label{flatf2}
ds_{\eta}^2 = \tilde{\eta}_{\mu\nu}d\tilde{x}^\mu d\tilde{x}^\nu   = -\frac{1}{(1-\xi)}
d\tilde{t}^2 + dr^2 + r^2 \{d\theta^2 + \sin^2\theta\; d\phi^2\}.
\end{equation}
Considering now
\begin{equation}\label{velocity2}
\tilde{V}^\mu = \gamma\, \left( \sqrt{1-\xi},\,-\beta^i\right);  \qquad \tilde{V}_\mu = -\gamma\, \left( \frac{1}{\sqrt{1-\xi}} ,\,\beta^i\right),
\end{equation}
with $\gamma$ given by equation (\ref{gammar}), and
\begin{equation}
\tilde{f}_{\mu\nu} = \tilde{\eta}_{\mu\nu} + \xi \tilde{V}_\mu \tilde{V}_\nu ,
\end{equation}
one can recover directly the foreground metric as given in (\ref{bhepsilon}). In this coordinate system the matrix 
$\gamma=\sqrt{g^{-1}f}$ and  its inverse are non-singular (in the matrix sense) as any horizon is approached. Thus once $L_\mathrm{int}$ 
is ``turned on'', the stress tensor $T^\mu{}_\nu$ is a non-singular matrix at the horizon, and can easily be
tuned to be perturbatively small. Turning on the massive graviton should then not qualitatively disturb the horizon structure. 

Thus we see that the generalized Gordon ansatz is sufficiently flexible to deal with Schwarzschild  
black holes (and Reissner--Nordstr\"om black holes).  This observation is important in that it gives us confidence 
that the ansatz is not too restrictive --- the physics results we are obtaining above really are due to the interaction 
Lagrangian, not due to an overly restrictive metric ansatz.

\section{Generalized Kerr--Schild ansatz}\label{s:ks}
\begin{figure}[htbp!]
\begin{center}
\includegraphics[scale=0.50]{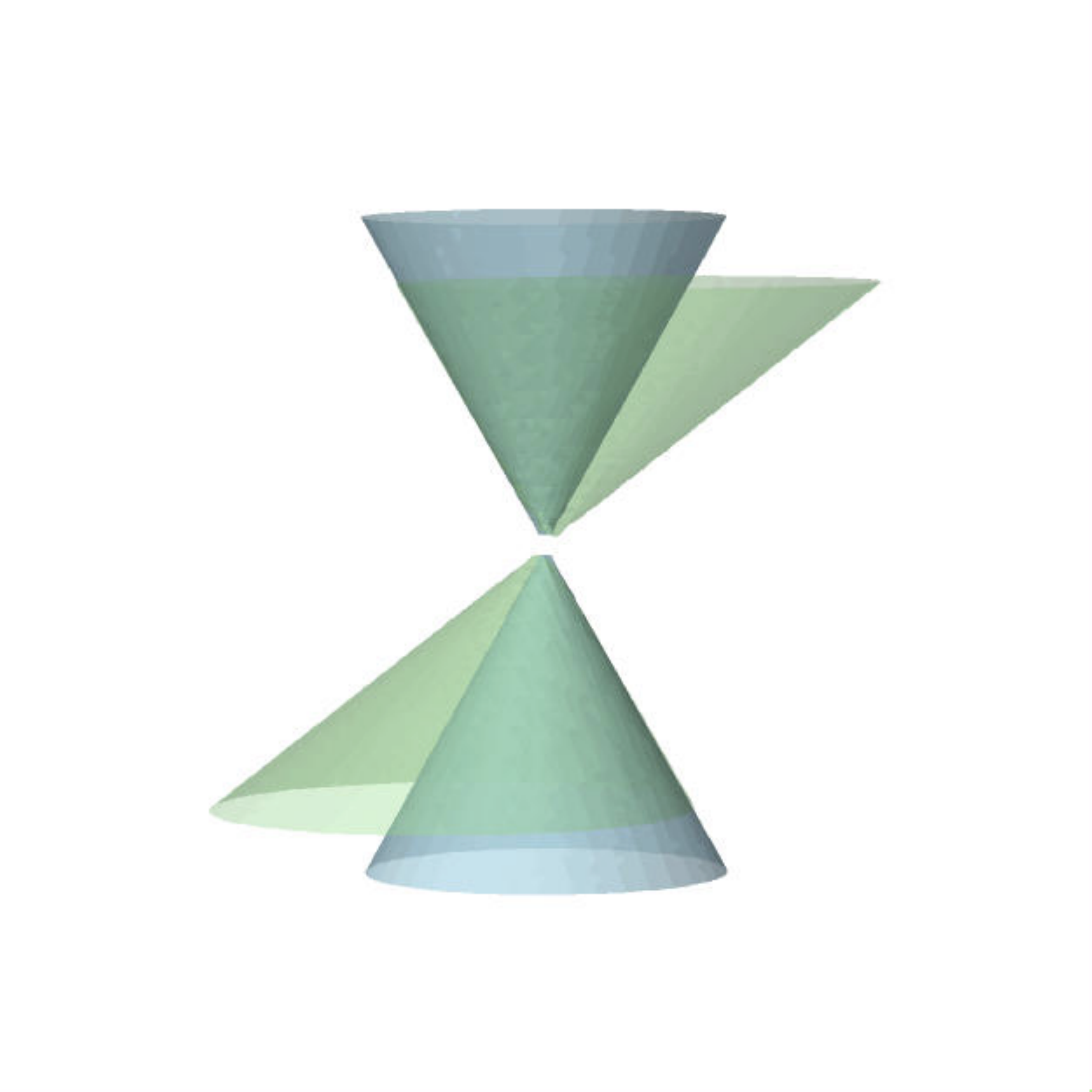}
\caption{Light cones for the generalized Kerr--Schild ansatz: Depending on the value of the parameter $\xi$, and apart from one null ray in common,  the foreground light cones lie strictly inside, on top of, or strictly outside the background light cones.}
\label{F:Kerr-Schild}
\end{center}
\end{figure}

We now consider an ansatz which can be thought as being inspired by the Kerr--Schild metrics~\cite{exact1, exact2}. This ansatz takes
a similar form to that considered in the previous section, namely
\begin{equation}\label{ansatz2}
f_{\mu\nu}=\Omega^2(g_{\mu\nu}+\xi \; l_\mu l_\nu )
\end{equation}
where the vector $l_\mu$ is now a 
null vector for foreground metric $g_{\mu\nu}l^\mu l^\nu=0$. So the geometries in question are conformally Kerr--Schild. It should be emphasized that the Kerr--Schild ansatz is an important and rather general one, which includes almost all spacetimes of physical interest~\cite{exact1, exact2}. 
Note that (see figure~\ref{F:Kerr-Schild}):
\begin{itemize}
\item For $\xi<0$, apart from a single null vector in common, the light cones of $f$ lie strictly outside the light cones of $g$.
\item For $\xi=0$ the light cones of $f$ and $g$ coincide; the two metrics are conformally related. 
(In this particular case the generalized Gordon and generalized Kerr--Schild ans\"atze coincide.)  
\item For $\xi>0$, apart from a single null vector in common, the light cones of $f$ lie strictly inside the light cones of $g$.
\end{itemize}
For this ansatz, one has
\begin{equation}\label{ansatz2met}
  g^{\mu\sigma}f_{\sigma\nu} =\Omega^2 \left\{ \delta^\mu{}_\nu +\xi \; l^\mu l_\nu  \right\}=\left(\gamma^2\right)^\mu{}_\nu,
\end{equation}
whence
\begin{equation}
 \gamma^\mu{}_\nu  = \Omega\left\{ \delta^\mu{}_\nu +\frac{1}{2}\xi\; l^\mu l_\nu  \right\},
\end{equation}
and in general
\begin{equation}
(\gamma^n)^\mu{}_\nu  = \Omega^n \left\{ \delta^\mu{}_\nu +\frac{n}{2}\;\xi\; l^\mu l_\nu  \right\}.
\end{equation}
The symmetric polynomials are given by
\begin{eqnarray}
 e_1(\gamma)=4\Omega; \qquad e_2(\gamma) = 6\Omega^2; \qquad e_3(\gamma) = 4 \Omega^3.
\end{eqnarray}
(Note these quantities are independent of $\xi$.) 
Thus, taking into account equations (\ref{stress}) and (\ref{stressT}), the stress-energy  tensor can in this situation be written  as
\begin{equation}
T^\mu{}_\nu = \mathcal{F} \, \, l^\mu l_\nu  +\Xi \, \delta^\mu{}_\nu ,
\end{equation}
where
\begin{equation}
 \mathcal{F}=-\frac{\Omega}{2} \xi \left(\alpha_1+2\Omega\alpha_2+\Omega^2\alpha_3 \right),
\end{equation}
and
\begin{equation}
 \Xi=-\Omega (3\alpha_1+3\Omega\alpha_2+\alpha_3\Omega^2).
\end{equation}
Note that this stress-energy  tensor can be interpreted as a null fluid (with null flux $\mathcal{F}$,  and pressure $\Xi$). This is a specialization of the ``type II'' stress-energy tensor considered by Hawking and Ellis~\cite{H&E}.  Satisfaction of the NEC in this situation is checked by contracting with a generic null vector (not necessarily $l^\mu$), and amounts to the condition $\mathcal{F}\geq0$.

 We can also write (\ref{stress})
\begin{equation}
\tau^\mu{}_\nu = \hat{\mathcal{F}} \; l^\mu l_\nu  +\hat\Xi \; \delta^\mu{}_\nu .
\end{equation}
By inspection
\begin{equation}
\hat{\mathcal{F}} = \mathcal{F}; \qquad  \hat \Xi = \Xi + L_\mathrm{int} = \Omega(\alpha_1+3\alpha_2\Omega+\alpha_3 \Omega^2).
\end{equation}
Finally,
\begin{equation}
\widebar T^\mu{}_\nu = \widebar{\mathcal{F}} \; l^\mu l_\nu  +\widebar\Xi \; \delta^\mu{}_\nu .
\end{equation}
Noting that now $\sqrt{-f} = \Omega^4 \sqrt{-g}$, we have
\begin{equation}
\widebar{\mathcal{F}}= -{1\over \Omega^4} \;\hat{\mathcal{F}}; \qquad  \widebar \Xi = -{1\over \Omega^4} \; \hat \Xi.
\end{equation}
Regarding the NEC the key observation is
\begin{equation}
\widebar{\mathcal{F}}= -{1\over \Omega^4} \; {\mathcal{F}}.
\end{equation}
We again very explicitly see how foreground and background NECs are anti-correlated~\cite{Baccetti:2012zz}.

\section{Discussion}\label{discussion}

In the technical analysis of ghost-free bimetric gravity a central role is played by the matrix square root $\gamma=\sqrt{g^{-1} f}$ relating the foreground and background metrics. Motivated by the fact that this object can be somewhat awkward to deal with, we have developed two lines of attack. In a companion paper we have developed a number of mathematical results that help us manipulate this quantity~\cite{Baccetti:2012zz}, ultimately leading to a rather general analysis of the interplay between the foreground and background NECs. In the current article we have taken a slightly different tack, and have characterized the circumstances under which it is possible to simply and explicitly compute this matrix square root. 

In order to simplify the problem we have assumed that the metrics are related in a very particular form, fulfilling what we
called the ``generalized Gordon ansatz''.  This ansatz can be motivated from the analogue spacetime programme, and has been proven
to be amazingly powerful in this paper. It implies that the effective stress-energy  tensor automatically takes the form of
a perfect fluid.  This ansatz is extremely natural in a cosmological setting, and helps one understand why bimetric FLRW cosmologies have proved to be relatively tractable. 

Moreover, it has also allowed us to recover easily the relation between the NECs in the two spaces~\cite{Baccetti:2012zz}.
That is, we have shown that the simultaneous fulfillment of the NEC associated to 
the modifications of general relativity in both foreground and background spaces is only possible if it saturates; 
that is if the effect of these modifications
are equivalent to the consideration of a cosmological constant. 

In counterpoint to  these rather strong conclusions, it should be noted that the generalized Gordon
ansatz seems sufficiently general to study the most common physical situations one might be interested in. In fact, we have shown that it can describe the general class of
cosmological solutions in bimetric gravity, as well as black holes in massive gravity.

We have also considered a less general ansatz based on the Kerr--Schild metrics. The effective stress-energy 
tensor in this case has not quite so natural a form; it is equivalent to a null fluid.
Nevertheless, even in this case similar conclusions can be extracted about the simultaneous fulfillment of both NECs --- this is
only possible in the trivial case in which both metrics have the same light cone structure.

\acknowledgments

VB acknowledges support by a Victoria University PhD scholarship.
PMM acknowledges financial support from the Spanish Ministry of
Education through a FECYT grant, via
the postdoctoral mobility contract EX2010-0854.
MV acknowledges support via the Marsden Fund and via a James Cook Research Fellowship, both administered by the Royal Society of New Zealand.


\end{document}